\documentclass[twocolumn,amsmath,amssymb,aps,pre,floatfix,nofootinbib]{revtex4}
\usepackage{graphicx,graphics,times}
\usepackage{bm}

\newcommand{\bra}[1]{\langle#1|}                    
\newcommand{\ket}[1]{|#1\rangle}                    
\newcommand{\OUTER}[2]{|#1 \rangle \langle #2|}

\begin{document}

\date{\today}

\title{Quantum mechanical description of measurement and the basic properties of state transformation }
\author{I. Marvian}
 \affiliation{Department of Physics,
Sharif University of Technology, Tehran, Iran\\iman@mehr.sharif.edu}
\begin{abstract}
In this paper, without any priori assumption about the
post-measurement state of system, we will examine how this state is
restricted by assuming each of these following assumptions. First,
by using this reasonable assumption that two successive measurements
should be describable as one measurement. Second, by assuming the
impossibility of faster than light signaling, "No-signaling
condition".  However, only by using these assumptions it is not
possible to obtain the usual projection postulate. Instead, by means
of a simple lemma, we will show that the density operator of system
after a measurement is a linear function of the density operator
determined by the usual post-measurement postulate. Furthermore we
will show this linear function has a Kraus representation. Finally,
we will discuss about the physical meaning of this consequence.
\end{abstract}


\maketitle
\section{INTRODUCTION}
A measurement on a system, most generally, is a process which
produces different macroscopic outcomes; the only restriction is
that initially measuring apparatus should have no correlations with
the system being measured which can affect the outcomes. Any process
with this property can be regarded as one measurement. Quantum
mechanics claims that for every measurement there exists a set of
positive operators like $\{F_\mu\},\sum_{\mu} F_\mu=I$, such that
the probability of occurrence of outcome $\mu$ is calculated by the
trace rule i.e. $p(\mu)=tr(\rho F_\mu )$, where $\rho$ is the
density operator of the system under consideration
\cite{nielchuang}. So, to calculate the probabilities we do not need
to know anything
more about the underlying mechanism of measurement.\\
On the other hand, a measurement in itself always contains different
time evolutions which finally correlate the state of the quantum
system with the value of a macroscopic classical variable. Hence the
need for consistency of quantum mechanical description of
measurement with the time evolution rules may impose some
constraints
on the possible time evolutions.\\
Usually to derive basic properties of quantum dynamics one considers
this reasonable assumption that two successive time evolution should
be describable as one time evolution \cite{sak}. Now similarly,
according to the concept of measurement, it seems reasonable to
assume a process including a time evolution which is followed by a
measurement, as one measurement; so the total process should be
describable as one measurement.\\
Let us first see how this property can be deduced from linearity of
time evolution. Suppose the state of a system after time evolution
is described by $E(\rho)$ where $E$ is a linear, positive and trace
preserving map. Generally, every linear map on the space of linear
operators can be represented by
\begin{equation}\label{34}
E(\rho)=\sum_{i} N_i \rho M_i,
\end{equation}
where $N_i$ and $M_i$ are suitable operators. For trace preserving
maps we have $\sum_{i} M_i N_i=I$. Suppose after this time evolution
we perform a measurement described by the set $\{F_\mu\}$ such that
$\sum_{\mu} F_\mu=I$. So the probability of outcome $\mu$ is
\begin{equation}\label{wrd}
p(\mu)=tr([\sum_{i} N_i \rho M_i]F_\mu )=tr(\rho [\sum_{i} M_i F_\mu
N_i]).
\end{equation}
Let $F^{'}_{\mu}$ be
\begin{equation}\label{newm}
F^{'}_\mu=\sum_{i} M_i F_\mu N_i.
\end{equation}
For an arbitrary $\rho$ the probability $p(\mu)$ is positive; so
according to  Eq. (\ref{wrd}), $F'_\mu$ is a positive operator; also
it is clear that $\sum_{\mu} F^{'}_\mu=I$ and $p(\mu)=tr(\rho
F^{'}_\mu)$. The probability of different outcomes in this process
are thus the same with a new measurement which is described by the
set $ \{F'_\mu\}$ \cite{note2}. For completeness, to show the whole
process is equivalent to one new measurement, we should show that
the state of system after the whole process is the same with the
post-measurement state of this new measurement. We postpone this
work until section (III).\\
Now imagine another process which includes two successive
measurements on the system. Someone can regard the outcomes of the
first measurement and the second ones together as the outcomes of
one new measurement \cite{note3}; so it seems reasonable assumption
that the total process should be describable as one measurement. It
is straightforward to see that because of the usual post-measurement
rule this property holds in quantum mechanics \cite{nielchuang}.\\
In this manner possible state transformations, including changes by
time evolutions or imposed by  measurements are such that these
reasonable properties hold, without any problem in definition of the
measurement process. On the other hand, for consistency of these
reasonable requirements with the trace rule, the possible state
transformations should satisfy some constraints. For example, if
cloning was possible, by cloning the state of  system and performing
measurement on the copies, one could perform measurements not
obeying the trace rule. But the impossibility of cloning is a
consequence of the linearity of time evolution \cite{wooters}. In
the present letter we will investigate these necessary
constraints  on the possible state transformation of system. \\
It can be easily shown that  nonlinear modifications of time
evolution, as given in \cite{Wein}, leads to the faster than light
communication \cite{Wein-Gis}. Furthermore assuming the
impossibility of faster than light signaling, "no-signaling
condition", one can obtain the basic properties of time evolutions,
namely linearity and complete positivity \cite{Gisin,Gisin2}.
Following this argument, we will also see that the post-measurement
state rule can be obtained using the "no-signaling condition." \\
Before following these ideas we should remark an important point
about the effects of initial correlations with the environment. In
the absence of initial correlations density operator of a system
after time evolution is a function of its present density operator
\cite{Preskill}. But in general the evolution of an open system may
be affected by its initial correlations with environment like
entanglement between the system and its environment or dependency of
the state of environment to the state of system. In this situation
the density operator of system after evolution is not necessarily a
function of its present density operator. \cite{Buzek} contains an
example of these cases which is corrected in \cite{Buze1}. As a
simple example suppose the interaction between a system an its
environment is governed by a unitary like $U$ such that
\begin{eqnarray*}\label{welct}
U\ \ket{\psi1}= \ket{1}_{sys}\ket{0}_{env},\ \ \ \ \ \ \  U\
\ket{\psi2}= \ket{0}_{sys}\ket{0}_{env},
\end{eqnarray*}
where
\[
\ket{\psi1}=\frac{\ket{1}_{sys}\ket{0}_{env}+\ket{0}{sys}\ket{1}_{env}}{\sqrt{2}}\]
and
\[
\ket{\psi2}=\frac{\ket{1}_{sys}\ket{0}_{env}-\ket{0}_{sys}\ket{1}_{env}}{\sqrt{2}}\]
For either of these states, the density operator of the system is
\[\rho_{sys}=\frac{\OUTER{0}{0}_{sys}+\OUTER{1}{1}_{sys}}{2}\] and the
density operator of environment is
\[\rho_{env}=\frac{\OUTER{0}{0}_{env}+\OUTER{1}{1}_{env}}{2}.\] After
the evolution, density operator of the system in two cases are
different. In the case of $\ket{\psi 1}$ density operator evolves to
$\OUTER{0}{0}_{sys}$, but in the case of $\ket{\psi 2}$ it evolves
to $\OUTER{1}{1}_{sys}$. Initially $\rho_{sys}$ and $\rho_{env}$ in
two cases is the same, and the only difference is the difference
between correlations. Hence it is clear that the effects of initial
correlation should arise in our argument; Furthermore we can deduce
that the linearity of time evolution and other dynamical
characteristics of quantum mechanics are not the immediate
consequence of its statics rules; Indeed they holds only in a
special condition.
\section{TIME EVOLUTION}
Imagine a time evolution of a system which may include interactions
with the environment. After this evolution we perform a measurement
on the system. As it has been already mentioned, we should be able
to assume the whole process as one measurement. When we perform a
measurement on the system the measurement apparatus should have no
such initial correlations correlations which can produce observable
effects. This is a necessary condition in every measurement. Hence
to regard the whole process as one measurement, it is necessary that
in the initial time evolution system have no effective initial
correlations with the environment; because in this picture the
environment can be regarded as a part of measuring apparatus. In
this manner this necessary
restriction on the time evolutions arise in our argument in a natural way.\\
Suppose with this time evolution $\ket{\psi}\bra{\psi}$ evolves to
$E(\ket{\psi}\bra{\psi})$. A priori we make no assumption about the
dynamics of pure states, for example it might be described by a
nonlinear equation. Let initially system be in
$\ket{\psi_i}\bra{\psi_i}$ with the probability $p_i$, so
$\rho_{in}=\sum_i p_i \ket{\psi_i}\bra{\psi_i}$, is the initial
density operator of the ensemble under consideration. After
evolution system is in $E(\ket{\psi_i}\bra{\psi_i})$ with the
probability $p_{i}$, thus it is described by $\rho^{'}_1=\sum_i p_i
E(\ket{\psi_i}\bra{\psi_i})$. Then we perform a measurement on the
system which is described by the set $\{F_{\mu}\}$. So the
probability of occurrence of outcome $\mu$ is $\sum_i p_i tr(
E(\ket{\psi_i}\bra{\psi_i})F_{\mu})$. But there should exist another
set of positive operators like $\{F^{'}_{\mu}\}, \sum_{\mu}
F^{'}_{\mu}=I$, which describes the whole process as one measurement
such that
 $p(\mu)=tr(\rho_{in} F^{'}_{\mu} )$. Hence
\begin{equation}\label{ded}
tr(\rho^{'}_1 F_{\mu})=tr(\rho_{in} F^{'}_{\mu} ).
\end{equation}
Consider another ensemble of pure states described by  $\rho_{in}$ ,
in which system is in $\ket{\phi_i}\bra{\phi_i}$ with the
probability of $q_i$ such that $\sum_i q_i
\ket{\phi_i}\bra{\phi_i}=\rho_{in} $. After the time evolution this
system is in $E(\ket{\phi_i}\bra{\phi_i})$ with the probability
$q_i$, so after evolution this system is described by
$\rho^{'}_2=\sum_i q_i E(\ket{\phi_i}\bra{\phi_i})$. Hence according
to Eq. (\ref{ded}) the probability of occurrence of outcome $\mu$ is
\begin{equation}\label{t}
tr(\rho^{'}_2F_\mu )=tr(\rho_{in}F^{'}_\mu ).
\end{equation}
Comparing with Eq. (\ref{ded}) shows
\begin{equation}\label{ct}
tr(\rho^{'}_{1}F_\mu )=tr(\rho^{'}_{2}F_\mu ).
\end{equation}
This equality should hold for any positive operator like $F_\mu$ ,
so we can deduce $\rho^{'}_1=\rho{'}_2$. Hence two systems which are
initially described by the same density operator, after evolution
still have the same density operator. Thus density operator of the
system after evolution, $\rho'$, can be expressed as a function of
its present density operator, $\rho_{in}$. We represent this fact by
$\rho'=E(\rho_{in})$. So we have
\begin{equation}\label{wedct}
tr(E(\rho)F_\mu )=tr(\rho_{in}F^{'}_\mu ).
\end{equation}
Hence
\begin{equation}\label{wedefgct} tr(E(p_1\rho_1+p_2\rho_2)F_\mu
)=tr( (p_1\rho_1+p_2\rho_2) F^{'}_\mu).
\end{equation}
But we know
\begin{eqnarray*}\label{wedw4t5efgct}
tr(E(\rho_1)F_\mu )= tr(\rho_{1}F^{'}_\mu ),\ \ \ tr(E(\rho_2)F_\mu
)= tr(\rho_{2}F^{'}_\mu ).
\end{eqnarray*}
 So
\begin{equation}\label{wedddfedct}
tr(E(p_1\rho_1+p_2\rho_2)F_\mu )=tr( p_1E(\rho_1)+p_2E(\rho_2)
F_\mu).
\end{equation}
Because this equality should holds for any positive operator like
$F_\mu$ we can deduce
\begin{equation}
E(p_1\rho_1+p_2\rho_2)=p_1E(\rho_1)+p_2E(\rho_2).
\end{equation}
Thus we conclude that time evolution is linear.\\
Note that we have made no specific assumption about the time
evolution, except about initial correlations.\\
With a similar argument which have been used in \cite{Gisin}, we
will show the complete positivity of time evolution. First note that
positivity and linearity does not imply complete positivity. For
example the map of $\rho$ to $\rho^{T}$ is a positive and linear map
but it is not completely positive \cite{Preskill}. In fact this
evolution can be implemented by a suitable interaction and initial
correlations between our system and its environment \cite{Buzek}.
For proving complete positivity we again need to assume that there
exists no effective initial correlation between our system and its
environment. Suppose there exists an imaginary ancillary system
which has no interaction with the outside.  We can use our argument
for the composite system, because it satisfies the necessary
assumptions; hence the evolution of the composite system should be
linear. Also we can obtain this result for both subsystems,  because
each subsystem has no interaction with the another. Suppose the
state of the composite system is a product state like
$\rho_{sys}\otimes\rho_{anc}$. It is obvious that the state of two
system after evolution should still remain a product state. So it
will be $E_{sys}\otimes E_{anc}(\rho_{sys}\otimes\rho_{anc})$. But
any state can be expanded in terms of product states. Hence for any
state of the composite system time evolution should be described by
$E_{sys}\otimes E_{anc}$. On the other hand the ancillary system has
no interaction with the outside, thus in principle its state can
remain constant. So, $E_{sys}\otimes I_{anc}$ is a legitimate
evolution of the composite system. Therefore for an arbitrary
density operator of the composite system, $\rho_{total}$,
$E_{sys}\otimes I_{anc}(\rho_{total})$ is a positive operator ,
whatever the dimension of the ancillary system. So, $E_{sys}$ is a
completely positive map and has a Kraus representation . On the
other hand, a linear, trace preserving and completely positive map
always can be realized quantum mechanically, with a unitary time
evolution on a larger Hilbert space \cite{Preskill}. So we have
obtained all the theoretical restrictions on an arbitrary time
evolution of system.
\section{POST-MEASUREMENT STATE}
Now by a similar argument like previous part we will derive  the
post-measurement state rule. Suppose by performing a measurement
which is described by the set $\{ F_{\mu}\}$, we observe result
$\mu$ with the probability $tr(\rho F_\mu)$ and the system jumps to
the state $\rho_\mu$. A priori we make no assumption about this
state; so it may be different for different ensembles of pure states
which are described initially by the same density operator. Now
performing another measurement which is described by the set
$\{G_{\nu}\}$, we obtain result $\nu$ with the probability
$tr(\rho_\mu G_\nu)$. So the probability of obtaining $\mu$ in the
first measurement and $\nu$ in the second one is
\begin{equation} p(\mu,\nu)=tr(\rho
F_\mu)\times tr(\rho_\mu G_\nu).
\end{equation}
On the other hand, as it has been assumed, we can regard the total
process as one measurement with outcomes $(\mu,\nu)$. So there
should exist a set of positive operators like $\{
\Omega_{\mu\nu}\},\sum_{\mu\nu}\Omega_{\mu\nu}=I$, such that
$p(\mu,\nu)=tr(\rho \Omega_{\mu\nu})$. So we can deduce
\begin{equation}\label{6klllta}
tr(\  [\rho_\mu tr(\rho F_\mu)]\  G_\nu)=tr(\rho\ \Omega_{\mu\nu}).
\end{equation}
Comparing this equation with Eq. (\ref{ded}), by the same argument
which we have used in that case, we can show that
$B_\mu(\rho)=\rho_\mu tr(\rho F_\mu)$ and  $\rho_{\mu}$ should be
the same for different ensemble described by one $\rho$ and moreover
$B_\mu(\rho)$ should be a linear function of $\rho$. Also by
definition it is clear that
\begin{equation}\label{6itdfddda}
tr(B_\mu(\rho))=tr(\rho F_\mu).
\end{equation}
In the appendix we will show that always there exists a linear,
positive, trace preserving map like $E_{\mu}$ such that
\begin{equation}\label{6ita}
B_\mu(\rho)=E_{\mu}(\sqrt{F_{\mu}}\rho \sqrt{F_{\mu}}).
\end{equation}
Hence according to the definition of $B_\mu$ the state of system
after obtaining result $\mu$ is
\begin{equation}\label{6ta}
A_\mu(\rho)=\frac{E_{\mu}(\sqrt{F_{\mu}}\rho
\sqrt{F_{\mu}})}{tr(\rho F_\mu)}.
\end{equation}
By a similar argument which we have used in the previous part, we
can see that in the absence of effective initial correlations, if
one consider an imaginary ancillary system, $B_{\mu}$ should be
replaced with $B_{\mu}\otimes I_{anc}$. So $B_\mu$ should be
completely positive and thus, as we will see in the appendix, we can
always choose $E_\mu$ to be completely positive. Therefore $E_\mu$,
which is a linear, trace preserving and completely positive map, can
be regarded as a time evolution of system for outcome $\mu$.
Regarding the concept of measurement this result seems reasonable,
because in a measurement different outcomes may have different time
evolutions. As we will see in the appendix, we can not specify
$E_\mu$ uniquely; but regarding Eq. (\ref{6ta}), this ambiguity has
no physical meaning. The necessity of quantum collapse can simply be
deduced from this post-measurement rule.\\
We call a measurement in which all $E_\mu$ are the identity maps an
\emph{ideal measurement}. Therefore we have shown that every real
measurement is equivalent to an ideal measurement which is followed
by different time evolutions for different outcomes. So
theoretically , all different manners of measuring one physical
property, are equivalent to the same ideal measurement
 followed by different time evolutions; these time evolution are dependent to the special
 manner of measuring that physical property. Note that in a
real measurement this two parts, ideal measurement and the following
time evolution, may be inseparable. For example in a Stern-Gerlach
measurement the outcome beams because of their different spins
obtain different phases in the magnetic field, which is equivalent
to a unitary time evolution. Actually, in more realistic model
different outcomes may experience different magnetic fields such
that they will be no longer described by mutually orthogonal state
vectors. As a better example suppose we are going to measure the
energy of an excited atom in the following manner. Returning an
excited electron to its ground state releases a photon; measuring
the frequency of this photon, we can measure the initial energy of
the excited atom. So we can regard this process as a measurement on
the atom. After measurement, the state of the atom is independent of
its initial state. So this process can be regarded as an ideal
energy measurement which projects the state of the atom to the
energy eigenstates followed by time evolutions which transform all
the energy eigenstates to the ground state. Note that regarding the
real process, in neither of these examples we cannot separate one
part of process as an ideal measurement and one part as
a time evolution; but, as we have shown, it is possible theoretically.\\
As an important special case, suppose the measurement process is
such that we can choose $E_\mu$ to be a unitary evolution specified
by a unitary operator like $U_\mu$. In this situation Eq.
(\ref{6ta}) becomes
\begin{equation}\label{e6ta}
A_\mu(\rho)=\frac{U_{\mu}(\sqrt{F_{\mu}}\rho
\sqrt{F_{\mu}})U_\mu^\dag}{tr(\rho F_\mu)}.
\end{equation}
According to the polar decomposition theorem \cite{nielchuang}, any
operator like $M_\mu$ can be decomposed to $M_\mu=V_\mu
\sqrt{M^\dag_\mu M_\mu}$, where $V_\mu$ is a unitary operator. So,
for a given set of operators like $\{M_\mu\}$ which satisfies
$\sum_\mu M^\dag_\mu M_\mu=I$, we can choose in Eq. (\ref{e6ta})
$U_\mu$ equal to $V_\mu$ and $F_\mu$ equal to $M^\dag_\mu M_\mu$. In
this manner we can deduce that, this set of operators describes a
measurement such that the probability of outcome $\mu$ in this
measurement is $tr(\rho\ M^\dag_\mu M_\mu)$ and the post-measurement
state of system is
\begin{equation}\label{dse6ta}
A_\mu(\rho)=\frac{M_{\mu} \rho M_\mu^\dag}{tr(\rho M^\dag_\mu
M_\mu)}.
\end{equation}
In fact this is the description of the well-known \emph{generalized
measurement} \cite{nielchuang}. But note that Eq. (\ref{6ta}) for
describing the post measurement state is more general and there
exists measurements which their post-measurement state does not obey
Eq. (\ref{dse6ta}). In the above example for measuring the energy of
an excited atom, it is straightforward to see that when the energy
levels have degeneracy, the post-measurement
state can not be described by Eq. (\ref{dse6ta}).\\
Now suppose the state of system after two successive measurement is
$A_{\mu\nu}(\rho)$. According to  Eq. (\ref{6klllta}), one can
easily show that $tr(\rho\ \Omega_{\mu\nu})A_{\mu\nu}(\rho)$ is a
linear, completely positive map. Also it is obvious that its trace
is equal to $tr(\rho\ \Omega_{\mu\nu})$. So according to the
appendix lemma, there exists always a completely positive and trace
preserving map like $E_{\mu\nu}$ such that
\begin{equation}\label{qlk}
A_{\mu\nu}(\rho)=\frac{E_{\mu\nu}(\sqrt{\Omega_{\mu\nu}} \rho
\sqrt{\Omega_{\mu\nu}})}{tr(\rho\Omega_{\mu\nu})},
\end{equation}
which clearly obeys the post-measurement rule Eq. (\ref{6ta}).
Therefore the total process is really equivalent with a measurement;
i.e. there exists a measurement such that it has the same outcome
state with the same probabilities. Because a time evolution can be
regarded as a special measurement with one outcome, so any sequence
of measurements and time evolution can be regarded as one
measurement.
\section{NO-SIGNALING CONDITION}
In a recent paper \cite{Gisin}, the authors by a simple argument
based on the impossibility of faster than light signaling, the
"no-signaling condition", have derived the linearity and complete
positivity of time evolution. Imagine an entangled pair which is
shared between Alice and Bob. From the no-signaling condition we can
deduce, performing a measurement on Alice's system does not affect
density operator of Bob's one. Furthermore using this condition,
without using the projection postulate, the authors have shown that
by performing suitable measurements on  Alice's system, one can
prepare all possible decomposition of the density operator of Bob's
system \cite{Gisin,Gisin2}. Now under an arbitrary time evolution of
Bob's system, all of these ensembles should remain
indistinguishable; otherwise this scheme can be used for faster than
light signaling. In this manner authors have shown that time
evolution should be linear. Also from the linearity and positivity
of time evolution they have deduced the complete  positivity. \\
Now with the similar argument which they have used for the time
evolution, we will show that our results about the post-measurement
rule can be derived using the no-signaling condition of our
assumption. We use this consequence of \cite{Gisin,Gisin2} that by
using the no-signaling condition and without using the projection
postulate, it can be shown that Alice by performing suitable
measurements can prepare every possible ensemble realization of the
density operator of Bob's system. Suppose one of these ensembles is
prepared. Now Bob performs a measurement which is described by $\{
F_{\mu}\}$ such that result $\mu$ is obtained with the probability
$tr(\rho F_\mu)$ and system jumps to the state $\rho_\mu$. A priori
we make no assumption about this state; so it may be different for
different ensembles of pure states which are described initially by
the same density operator. Now Bob performs another measurement
which is described by $\{G_{\nu}\}$ and obtains result $\nu$ with
the probability $tr(\rho_\mu G_\nu)$; But we know that this
probability should be the same for different ensembles of pure
states  which are described by one density operator; otherwise, if
it was different from one to another, Alice by performing different
measurements on her own system and preparing different
decompositions of the density operator of Bob's system could send
faster than light signals. But here $G_\nu$ is an arbitrary positive
operator; hence $\rho_\mu$ should be the same for different
preparations of a density operator and can be expressed as
a function of $\rho$, like $A_\mu(\rho)$.\\
Now suppose the system is in $\rho_1$ with the probability $p_1$ and
in
 $\rho_2$ with the probability $p_2$. Bob performs a measurement
which is described by $\{F_\mu \}$. For the state $\rho_1$ result
$\mu$ is obtained with the probability $tr(\rho_1 F_\mu)$ and the
system jumps to $A_\mu(\rho_1)$; for the state $\rho_2$ result $\mu$
is obtained with the probability $tr(\rho_2 F_\mu)$ and then the
system jumps to $A_\mu(\rho_2)$. So, for the ensemble under
consideration, we obtain result $\mu$ with the probability $p_1
tr(\rho_1 F_\mu)+p_2 tr(\rho_2 F_\mu)$ and after obtaining this
result the state of system is
\begin{equation}\label{gfta}
\frac{p_{1} tr(\rho_{1} F_{\mu}) A_\mu(\rho_{1})}{p_{1} tr(\rho_{1}
F_{\mu})+p_{2} tr(\rho_{2} F_{\mu})}+\frac{p_{2} tr(\rho_{2}
F_{\mu}) A_\mu(\rho_{2})}{p_{1} tr(\rho_{1} F_{\mu})+p_{2}
tr(\rho_{2} F_{\mu})}.
\end{equation}
On the other hand, initially the system is described by
$p_1\rho_1+p_2\rho_2$, hence after obtaining result $\mu$ the system
should jump to $A_\mu(p_1\rho_1+p_2\rho_2)$. Equating
$A_\mu(p_1\rho_1+p_2\rho_2)$ with expression (\ref{gfta}), one can
easily show that $tr(\rho F_\mu)A_\mu(\rho)$ is a linear function of
$\rho$. Now we can easily follow all the arguments which results
Eqs. (\ref{6ita},\ref{6ta}).
\section{DISCUSSION AND CONCLUSION}
As we have already mentioned, most generally any process which
produces different outcomes is a measurement on the system; the only
necessary condition is the lack of effective initial correlations.
For example consider an experimentalist who performs different
measurements and then after his observations produce an outcome. Now
the whole instruments and the experimentalist altogether can be
regarded as the measuring apparatus and the whole process can be
regarded as one measurement. The only condition is that there should
exist no initial correlation which can affect the outcomes. For
example the initial state of the system should be necessarily
unknown to the experimentalist. On the other hand, if he know
anything about the state of system, the total process can not be
regarded as one measurement. Indeed knowing this information, he can
produce outcomes such that their probabilities do not obey
the usual trace rule.\\
We have seen that the quantum mechanical description of measurement
and especially the rule for the outcome's probabilities in quantum
mechanics are such that this property holds in the theory if and
only if the outcomes of a measurement be described by Eq.
(\ref{6ta}). This has been driven with the help of a simple lemma.
The necessity of quantum collapse is a consequence of this equation.
As we have seen, theoretically all different manners of measuring
one physical property are equivalent with a special \emph{ideal
measurement}, associated to the physical property, followed by time
evolutions which depends to the special manner of measuring. Also we
have seen how the description of the generalized measurements can be
obtained as a special case. We have mentioned an example which its
post-measurement state cannot be described by the post-measurement
state rule of generalized measurements, but can be described by Eq. (\ref{6ta}).\\
At the end we have seen that from the impossibility of faster than
light signaling one can also derive this post-measurement state
rule. In this manner we have completed the main purpose of
\cite{Gisin} to derive  fundamental properties of quantum
transformations from the no-signaling condition and the usual trace
rule.
\section{Appendix}
\emph{Lemma}: Let $B$ be a linear and positive map on the space of
linear operator which satisfies
\begin{equation}\label{srewe}
 tr(B(\rho))=tr(\rho F),
\end{equation}
where $F$ is a positive operator. Then there exists a linear,
positive and trace preserving map like $E$ such that
\begin{equation}\label{srlk}
B(\rho)=E(\sqrt{F}\rho \sqrt{F}).
\end{equation}
Furthermore for a
completely positive $B$, always $E$ can be chosen completely positive.  \\
\indent \emph{Proof}: Suppose $\{\ket{k}_I\}$ are the eigenstates of
$F$ with nonzero eigenvalues and $P_I$ is the projective operator to
this subspace. Also suppose $\{\ket{l}_{II}\}$ are the eigenstates
with zero eigenvalues and $P_{II}=I-P_I$ is the projective to this
subspace. It is obvious that
\begin{equation}\label{hhhh6ita}
B(\rho)=B(P_{I}\rho P_{I})+B(P_{I}\rho P_{II}+P_{II}\rho
P_{I})+B(P_{II}\rho P_{II}).
\end{equation}
$B$  is a positive map so the last term in Eq. (\ref{hhhh6ita})
should be a positive operator. Because $tr(B(\rho))=tr(\rho F)$,
this term is traceless; so it is always zero. Now we will show the
second term in the right-hand side is also zero. Consider a  vector
like $\ket{\psi}=\alpha\ket{k}_I+\beta \ket{l}_{II}$ with real
$\alpha,\beta$. In this situation
\begin{equation}
\label{awe} B(\ket{\psi}\bra{\psi})=\alpha^2\ B(\ket{k}_I\bra{k}_I)+
\alpha\beta\ B(\ket{k}_I\bra{l}_{II}+\ket{l}_{II}\bra{k}_{I}).
\end{equation}
The left-hand side should be positive. In the other side although
$B(\ket{k}_I\bra{k}_I)$ is positive, but the second term is not
necessarily so. Thus there exists $\alpha,\beta$ which makes
right-hand side non positive, in contradiction with the left-hand
side. So any term like
$B(\ket{k}_I\bra{l}_{II}+\ket{l}_{II}\bra{k}_{I}) $ should be zero.
Also one can repeat such an argument for another kind of terms like
$B(i\ket{k}_I\bra{l}_{II}-i\ket{l}_{II}\bra{k}_{I}) $ with the same
result. Because the second term in the right-hand side of Eq.
(\ref{hhhh6ita}) is just a linear combination of these two kind of
terms, it will vanishes; so we can conclude that
\begin{equation}\label{wwa}
B(\rho)=B(P_{I}\rho P_{I}).
\end{equation}
Let $F^{-1}$ be an operator which satisfies $F^{-1} F=P_I$. Now
suppose an arbitrary positive, trace preserving map like $E^{'}$. We
define $E$ to be
\begin{equation}\label{weedgwa}
E(\rho)=B(\sqrt{F^{-1}}P_I \rho P_I\sqrt{F^{-1}})+E^{'}(P_{II}\rho
P_{II}).
\end{equation}
Clearly it is a positive map; regarding Eq. (\ref{srewe}), it is
trace preserving. Also according to Eq. (\ref{wwa}) it obviously
satisfies
Eq. (\ref{srlk}).\\
By choosing $E^{'}$ to be completely positive, from complete
positivity of $B$ one can deduce complete positivity
of $E$. \\
\section{ACKNOWLEDGMENT}
I would like to thank Prof. S. Hosseini Khayat for his valuable
remarks.
{}


\begin{thebibliography}{}
\bibitem{note3} More generally, any sequence of different time evolutions
 and measurements should be describable as one measurement.
\end{thebibliography}
\end{document}